\documentclass[aps,prd,superscriptaddress,11pt,showpacs,notitlepage]{revtex4}
	\usepackage{graphicx}
	\usepackage{amsmath,amssymb,mathrsfs}
	\usepackage[colorlinks=true, a4paper=true, pdfstartview=FitV,
linkcolor=blue, citecolor=blue, urlcolor=blue]{hyperref}

\usepackage{epsf}
\usepackage{epsfig}

\newcommand{\be}{\begin{equation}}
\newcommand{\ee}{\end{equation}}
\newcommand{\bea}{\begin{eqnarray}}
\newcommand{\eea}{\end{eqnarray}}

\newcommand{\bb}{\bibitem}
 
\newcommand{\eqn}{\begin{eqnarray}}
\newcommand{\eqnx}{\end{eqnarray}}

\numberwithin{equation}{section}

\begin{document}
\title{BPS submodels of the Skyrme model}

\author{C. Adam}
\affiliation{Departamento de F\'isica de Part\'iculas, Universidad de Santiago de Compostela and Instituto Galego de F\'isica de Altas Enerxias (IGFAE) E-15782 Santiago de Compostela, Spain}
\author{J. Sanchez-Guillen}
\affiliation{Departamento de F\'isica de Part\'iculas, Universidad de Santiago de Compostela and Instituto Galego de F\'isica de Altas Enerxias (IGFAE) E-15782 Santiago de Compostela, Spain}
\author{A. Wereszczynski}
\affiliation{Institute of Physics,  Jagiellonian University,
Lojasiewicza 11, Krak\'{o}w, Poland}

\begin{abstract}
We show that the standard Skyrme model without pion mass term can be expressed as a sum of two BPS submodels, i.e., of two models whose static field equations, independently, can be reduced to first order equations. Further, these first order (BPS) equations have nontrivial solutions, at least locally. These two submodels, however, cannot have common solutions. Our findings also shed some light on the rational map approximation.
Finally, we consider certain generalisations of the BPS submodels.
\end{abstract}

\maketitle 
\section{Introduction}
The Skyrme model \cite{skyrme} is a nonlinear field theory supporting topological soliton solutions ("Skyrmions"), which finds its main application as a low-energy effective theory for QCD \cite{nappi}. In the analysis of Skyrme models (and related theories supporting topological solitons), two important concepts are topological energy bounds and the related notion of  Bogomolnyi equations \cite{Bogom1976}, \cite{Prasad1975}. Indeed, sometimes it is possible to reduce the static field equations to first-order equations ("Bogomolnyi equations") such that the corresponding solutions saturate the bound. Within the set of generalised Skyrme models (but all based on the same Skyrme field $U \in$ SU(2)), two cases of BPS Skyrme models satisfying a Bogomolnyi equation are known. The first one is the "BPS Skyrme model" consisting of a term sextic in first derivatives and a potential \cite{BPS}. In this model, Bogomolnyi solutions exist for arbitrary topological degree ("baryon number") $B$. The second one consists of the Skyrme term (quartic in first derivatives) and a particular potential \cite{Derek}. In this second case, however, only the $|B|=1$ solutions (and, of course, the trivial vacuum solution) saturate the bound and obey the corresponding Bogomolnyi equation. Higher $B$ configurations are unbound in this model and turn into lightly bound Skyrmions once further terms are added with sufficiently small coupling constants \cite{Martin}, \cite{Bjarke}, \cite{Bjarke2}. Both these BPS models are genuine Skyrme models in the sense that they can be found within the set of generalised Skyrme model by an appropriate choice of coupling constants. 

In the present paper we want to consider a slightly different type of BPS submodels. We will find that the standard Skyrme model (consisting of the sigma model term and the Skyrme term), too, may be written as a sum of two BPS submodels, where each of the two submodels separately leads to a Bogomolnyi equation and nontrivial solutions (at least, locally). These submodels, however, are not Skyrme models on their own, i.e., it is not possible to get just one of these submodels by a choice of coupling constants within the set of generalised Skyrme models. In other words, the two submodels are multiplied by the same coupling constants, and eliminating one automatically eliminates the other. Further, the two submodels have no common solutions (except for the trivial vacuum solution).  This must, of course, be expected, because it is known that, although the standard Skyrme model has a topological energy bound (the Skyrme-Faddeev bound \cite{skyrme}, \cite{faddeev}) and a corresponding Bogomolnyi equation, this first-order equation is too restrictive and only allows for the trivial solution. Despite these impediments, nevertheless, the existence of further BPS submodels is of interest and sheds new light on several issues within the Skyrme model.
    
The generalised Skyrme model we want to consider consists of four terms. The sigma model (Dirichlet) term 
\begin{equation}
\mathcal{L}_2=\frac{1}{2} \mbox{Tr}\; (L_\mu L^\mu),
\end{equation}
the Skyrme (quartic) term
\begin{equation}
 \mathcal{L}_4=\frac{1}{16} \mbox{Tr} \; ([L_\mu , L_\nu]^2), 
\end{equation}
and the sextic term, which is just the baryonic current squared term
\begin{equation}
\mathcal{L}_6=\lambda^2 \pi^4  \mathcal{B}_\mu \mathcal{B}^\mu, \;\;\; \mathcal{B}^\mu = \frac{1}{24\pi^2} \epsilon^{\mu \nu \rho \sigma} \mbox{Tr} \; L_\nu L_\rho L_\sigma, 
\end{equation}
where $ L_\mu \equiv U^\dagger \partial_\mu U$ is the left invariant current. The last term, without derivatives, is the potential $\mathcal{L}_0(U) = - \mu^2 \mathcal{U}({\rm tr} \, U)$ which provides a mass for pionic excitations. 
Here, we assume that units of energy and length have been chosen such that the coupling constants of the Dirichlet and Skyrme term are scaled away (Skyrme units). $\lambda$ and $\mu$ are, thus, dimensionless coupling constants controlling the relative strengths of the corresponding terms.

The original matrix $SU(2)$ Skyrme field $U$ can be decomposed into a real scalar $\xi$ and complex scalar $u$ by 
\begin{equation}
U= \exp (i\xi \vec\tau \cdot \vec n) = \cos \xi \, {\bf 1} + i \sin \xi \, \vec \tau \cdot \vec n
\end{equation}
\be
\vec{n}=\frac{1}{1+|u|^2} \left( (u+u^*), -i(u-u^*), 1-|u|^2 \right) .
\ee
For topologically nontrivial configurations the full target space $\mathbb{S}^3$ has to be covered, which means that $\xi $ should take values in the full interval $[0,\pi]$ and $u$ should take values in the full complex plane $ \mathbb{C}$. 
Each term of the generalized Skyrme model can then be rewritten in this new target space coordinates as
\be
\mathcal{L}_2 = \mathcal{L}_2^{(1)}+\mathcal{L}_2^{(2)}\; , \quad  \mathcal{L}_2^{(1)} \equiv 4\sin^2\xi \frac{u_\mu \bar{u}^\mu}{(1+|u|^2)^2}
\; , \quad \mathcal{L}_2^{(2)} \equiv \xi_\mu \xi^\mu
\ee
\bea
\mathcal{L}_4 =  \mathcal{L}_4^{(1)}+\mathcal{L}_4^{(2)} \; , \quad 
 \mathcal{L}_4^{(1)} &\equiv & 4 \sin^2 \xi \left( \xi_\mu \xi^\mu \frac{u_\mu \bar{u}^\mu}{(1+|u|^2)^2} -\frac{\xi_\mu \bar{u}^\mu \;\xi_\mu u^\nu}{(1+|u|^2)^2} \right) \nonumber \\
\mathcal{L}_4^{(2)} &\equiv & 4\sin^4 \xi  \frac{(u_\mu \bar{u}^\mu)^2-u_\mu^2 \bar{u}^2_\nu }{(1+|u|^2)^4}
\eea
\be
\mathcal{L}_6=\frac{  \lambda^2 \sin^4 \xi}{(1+|u|^2)^4} \;\left( i \epsilon^{\mu 
\nu \rho \sigma} \xi_{\nu} u_{\rho} \bar{u}_{\sigma} \right)^2\; , \quad \mathcal{L}_0 = -\mu^2 \mathcal{U}(\xi).
\ee
For later convenience we have divided the quadratic and quartic terms into two parts. Obviously, within the Skyrme model context, it is not possible to eliminate, e.g., $\mathcal{L}_2^{(1)}$ without eliminating, at the same time, also $\mathcal{L}_2^{(2)}$.

A well-known example of a Skyrme theory with the BPS property is the BPS Skyrme model \cite{BPS}, and it will be useful to review it briefly. It is defined as
\be
\mathcal{L}_{BPS}= \mathcal{L}_6 + \mathcal{L}_0
\ee
which after the field decomposition reads
\be
\mathcal{L}_{BPS}=\lambda^2 \frac{ \sin^4 \xi}{(1+|u|^2)^4} \;\left( i \epsilon^{\mu 
\nu \rho \sigma} \xi_{\nu} u_{\rho} \bar{u}_{\sigma} \right)^2 - \mu^2 \mathcal{U}(\xi ) .
\ee
The Bogomolny equation for static field configurations is
\be \label{BPS-bog}
\lambda \frac{ \sin^2 \xi}{(1+|u|^2)^2} \; i \epsilon^{ijk} \xi_{i} u_{j} \bar{u}_{k} =\pm   \mu \sqrt{\mathcal{U}} .
\ee
It is straightforward to notice that there are other possibilities to distribute the derivatives of the fields and their contractions with the epsilon symbol to form new first order equations with the same field content. Therefore, they can be treated as new Bogomolnyi equations for the pertinent Skyrme-like models. This will be analyzed in the rest of the paper.   

\section{The first BPS submodel}
$\mathcal{L}_2^{(1)}$ and $\mathcal{L}_4^{(1)}$ can be combined into the following expression
\be
\mathcal{L}^{(1)}=4 \sin^2 \xi  \left( \frac{u_\mu \bar{u}^\mu}{(1+|u|^2)^2}+ \frac{ (\xi_\mu u^\mu)(\xi_\nu \bar{u}^\nu)  - \xi_\mu^2 (u_\nu\bar{u}^\nu)}{(1+|u|^2)^2} \right) .
\ee
The energy of the static case reads 
\bea
{E}^{(1)}&=& 4 \int d^3 x \frac{\sin^2 \xi}{ (1+|u|^2)^2} \left[ u_i  \bar{u}_i +  (i\epsilon_{ijk} \xi_j u_k) (-i\epsilon_{imn} \xi_m \bar{u}_n)  \right] \\
&=& 4\int d^3 x \frac{\sin^2 \xi}{(1+|u|^2)^2} (u_i\pm i \epsilon_{ijk} \xi_j u_k)(\bar{u}_i \mp i \epsilon_{imn} \xi_m \bar{u}_n)  \mp 8 \int d^3 x  \frac{i\sin^2 \xi}{(1+|u|^2)^2} \epsilon_{ijk} \xi_i u_j \bar{u}_k \nonumber \\
&\geq & 8 \left| \int d^3 x  \frac{i\sin^2 \xi}{(1+|u|^2)^2} \epsilon_{ijk}  \xi_i u_j\bar{u}_k \right| 
= 8\pi^2 \left| \int d^3x  B_0 \right| = 8\pi^2 |B|  \nonumber
\eea
where the bound is saturated for solutions of the following Bogomolnyi equation
\be
u_i\pm i \epsilon_{ijk} \xi_j u_k=0
\ee
and its complex conjugation. Note that this equation implies some constraints for the fields, namely 
\be \label{constr-1}
u_i \xi_i=\bar{u}_i \xi_i=0, \;\;\; u_i^2=\bar{u}_j^2=0 .
\ee
It is interesting to consider the particular solutions provided by the ansatz in spherical polar coordinates $\xi = \xi (r)$, $u=u(\theta , \varphi)$.
This ansatz automatically satisfies the first constraint $u_i \xi_i =0$. To simplify the second constraint, it is useful to use the stereographic projection from the unit sphere spanned by $(\theta ,\varphi )$ to the plane spanned by $(x,y)$, say, where $z\equiv x+iy = \tan (\theta/2) e^{i\varphi}$, because the metric on the unit sphere is conformally flat in the coordinates $(x,y)$, such that the constraint $u_i^2 =0$ simplifies to $u_z u_{\bar{z}}=0$, i.e., $u$ must be either holomorphic or anti-holomorphic in the complex coordinate $z$. The sign choice in the Bogomolnyi equation (together with the boundary conditions imposed on $\xi$) determine whether $u$ is holomorphic or anti-holomorphic.

It is instructive to insert the ansatz in spherical polar coordinates directly into the energy functional. In a first step we get $\epsilon_{ijk} \xi_j u_k \epsilon_{imn} \xi_m \bar u_n = \xi_j^2 u_i \bar u_i$, such that both terms are proportional to $u_i \bar u_i$. Next, using the metric in these coordinates
\be
ds^2 = dr^2 + r^2 ds_{\mathbb{S}^2}^2 \; , \quad ds_{\mathbb{S}^2}^2 = \frac{4}{1+x^2 + y^2}(dx^2 + dy^2)
\ee
we get for the volume element
\be
d\Omega_{\mathbb{R}^3} = dr r^2 d\Omega_{\mathbb{S}^2} \; , \quad d\Omega_{\mathbb{S}^2} = \frac{2i}{(1+z\bar z)^2} dz d\bar z
\ee 
and, further,
\be
u_i \bar u_i \equiv g^{ij}u_i \bar u_j = \frac{(1+z\bar z)^2}{2r^2} (u_z \bar u_{\bar z} + u_{\bar z}\bar u_z).
\ee
As a result, the energy functional factorises,
\be
E^{(1)} = 2E^{(1)}_\xi  E^{(1)}_u
\ee
where
\be
E^{(1)}_u = \int d\Omega_{\mathbb{S}^2} \frac{(1+z\bar z)^2}{(1+u\bar u)^2} (u_z \bar u_{\bar z} + u_{\bar z} \bar u_z)
\ee
is just the CP(1) (non-linear sigma) model on $\mathbb{S}^2$. Its finite energy solutions are provided by all holomorphic (positive winding number) and anti-holomorphic (negative winding number) rational functions. These solutions saturate the Bogomolnyi bound $E^{(1)}_u \ge 4\pi |N|$, where the winding number $N$ is given by the degree of the rational map.

The second energy functional is
\be
E^{(1)}_\xi = \int dr \sin^2 \xi (1+\xi'^2)
\ee
and allows for an almost trivial Bogomolnyi bound,
\be
  E^{(1)}_\xi = \int dr \sin^2 \xi \left( (1\mp \xi')^2 \mp 2 \xi' \right) \ge 2\left| \int dr \sin^2\xi \xi' \right| = 2 \int_0^\pi \sin^2 \xi d\xi =\pi 
\ee
where we used the boundary conditions $\xi (0)=\pi$, $\xi (\infty )=0$.  The corresponding Bogomolnyi equation is 
\be
\sin^2 \xi (\xi'^2\pm 1)^2 =0,
\ee 
and the solution with the right boundary conditions is the compacton
\be
 \xi (r) =
\left\{
\begin{array}{c}
\pi -r \quad \mbox{for} \quad 0\le r \le \pi \\
0 \quad \mbox{for} \quad  r>\pi .
\end{array}
\right.\, 
\ee
The first derivative of the compacton is not continuous at the compacton boundary $r=\pi$, but the energy density is continuous, and both the Bogomolnyi equation and the full second-order Euler-Lagrange (EL) equation hold everywhere in space for this solution, owing to the presence of the factor $\sin^2 \xi$.
It may be checked easily that the baryon number $B$ is equal to the winding number $N$ for this ansatz. Finally, the energy is $E^{(1)} = 2 \cdot \pi \cdot 4\pi |N| = 8\pi^2 |B|$, as it must be.

In particular, it follows that this submodel has BPS solutions for arbitrary rational maps $u=R(z) = p(z)/q(z)$ (where $p$ and $q$ are polynomials without common divisor). This is interesting, because rational maps have been employed to construct approximate solutions for the full standard Skyrme model $\mathcal{L}_2 + \mathcal{L}_4$ \cite{rat-map}.  Rational maps cannot be genuine solutions of this model, because the rational map ansatz is incompatible with the EL equation resulting from the term  $\mathcal{L}_4^{(2)}$ (except for $|B|\le 1$). Inserting the rational map ansatz directly into the corresponding energy functional, nevertheless, defines a {\em restricted} variational problem. The energy now depends on the particular rational map, and minimisation leads to rational maps with interesting discrete symmetries (e.g. the symmetries of platonic solids) for $|B|\ge 3$. Further, these symmetries agree with the symmetries of the full numerical solutions \cite{shells}, and also the energies of the corresponding rational map approximations are rather close to the energies of the numerically calculated Skyrmions. Here we may conclude that the rather good quality of the rational map approximation may be understood from the fact the model has a BPS submodel which is exactly solved by rational maps, and there is only one term ($\mathcal{L}_4^{(2)}$) which prevents rational maps from being exact solutions.

\section{The second BPS submodel}
$\mathcal{L}_2^{(2)}$ and $\mathcal{L}_4^{(2)}$ can be combined into the expression
\be
\mathcal{L}^{(2)}= \xi_\mu \xi^\mu +4\sin^4 \xi  \frac{(u_\mu \bar{u}^\mu)^2-u_\mu^2 \bar{u}^2_\nu }{(1+|u|^2)^4}  .
\ee
The energy functional for static configurations reads
\bea
E^{(2)}&=& \int d^3 x \left( \xi_i^2 + 4\sin^4 \xi \frac{1}{(1+|u|^2)^4} (i\epsilon_{ijk} u_j \bar{u}_k)^2 \right) \\
&= & \int d^3 x  \left(\xi_i \mp  \frac{2i\sin^2 \xi}{(1+|u|)^2} \epsilon_{ijk} u_j \bar{u}_k \right)^2 \pm 4\int d^3x  \frac{i\sin^2 \xi}{(1+|u|)^2} \epsilon_{ijk} \xi_i u_j \bar{u}_k  \nonumber \\
&\geq &  4 \left| \int d^3x  \frac{i\sin^2 \xi}{(1+|u|)^2} \epsilon_{ijk} \xi_i u_j \bar{u}_k \right| 
= 4\pi^2 \left| \int d^3x  B_0 \right| = 4\pi^2 |B| .  \nonumber
\eea
The bound is saturated for solutions of the Bogomolnyi equations
\be \label{(2)-bog}
\xi_i \mp  \frac{2i\sin^2 \xi}{(1+|u|)^2} \epsilon_{ijk} u_j \bar{u}_k =0 .
\ee
 Observe  that the Bogomolnyi equations lead to some constrains for the fields,
\be \label{constr-2}
u_i\xi_i = \bar{u}_i \xi_i=0.
\ee
Assuming the ansatz $\xi = \xi (r)$, $u=g(\theta) e^{im\varphi}$, we get for $u$ the solution
\be \label{con-sing}
u= \tan \frac{\theta}{2} e^{im\varphi},
\ee
which, for $|m|>1$, has a conical singularity along the $z$ axis. The energy density and winding number density, on the other hand, are smooth.
Further, $m$ is equal to the baryon number, $B=m$, for genuine Skyrmion configurations (i.e., where $\xi$ obeys the corresponding boundary conditions). 
It is interesting to note that the BPS Skyrme model leads to the same solution for $u$ this ansatz. The two BPS equations (\ref{BPS-bog}) and (\ref{(2)-bog}) are, in fact, very similar for this ansatz (identical for $u$, different for $\xi (r)$).
The resulting equation for $\xi$ reads
\be
\frac{d\xi}{dr}=\pm \frac{m}{r^2}{\sin^2 \xi}.
\ee
Choosing the minus sign (a negative slope for $\xi (r)$), the
 solution is
\be
\xi = \mbox{ arccot } m \left( s_0 - \frac{1}{r}  \right) 
\ee
where $s_0$ is an integration constant. At $r\to 0$ we have $\cot \xi \to -\infty$, i.e., $\xi \to \pi$, as desired. In the limit $r\to \infty$, however, $\cot \xi$ does not approach $\infty$, i.e., $\xi $ does not approach 0. Instead, the profile function $\xi$ takes values only in the interval $\pi \ge \xi (r) \ge \mbox{arccot}\, ms_0 >0$. The local solution of the BPS equation, therefore, cannot be extended to a solution on the full target space, i.e., to a Skyrmion. Instead, it leads to a "fractional" Skyrmion, where both the baryon number and the BPS energy may take arbitrary fractional values, defined by the choice of the integration constant $s_0$.

\section{Some generalisations}
We now want to consider some generalisations of the two BPS submodels, by multiplying each term by a certain coupling function.
\subsection{The Dilaton-YM like model}
We consider the following model 
\be
\tilde{\mathcal{L}}^{(2)}=h_2 (\xi, u\bar{u}) \left( \xi_\mu \xi^\mu +g^2_2 (\xi, u\bar{u})\sin^4 \xi  \frac{(u_\mu \bar{u}^\mu)^2-u_\mu^2 \bar{u}^2_\nu }{(1+|u|^2)^4} \right) .
\ee
The energy integral for static configurations reads
\bea
\tilde E^{(2)} &=& \int d^3 x h_2 \left( \xi_i^2 +g^2_2 \sin^4 \xi \frac{1}{(1+|u|^2)^4} (i\epsilon_{ijk} u_j \bar{u}_k)^2 \right) \\
&= & \int d^3 x h_2 \left(\xi_i \mp  \frac{ig_2 \sin^2 \xi}{(1+|u|)^2} \epsilon_{ijk} u_j \bar{u}_k \right)^2 \pm 2\int d^3x  \frac{ig_2h_2\sin^2 \xi}{(1+|u|)^2} \epsilon_{ijk} \xi_i u_j \bar{u}_k  \nonumber \\
&\geq &  2 \left| \int d^3x  \frac{ig_2h_2\sin^2 \xi}{(1+|u|)^2} \epsilon_{ijk} \xi_i u_j \bar{u}_k \right| \nonumber \\
&=& 2\pi^2 \left| \int d^3x g_2h_2 B_0 \right| = 2\pi^2 |B| \langle g_2h_2 \rangle _{\mathbb{S}^3} \nonumber
\eea
where $\langle \cdot \rangle_{\mathbb{S}^3} $ is the target space average of the target space function inserted between the brackets.
The bound is saturated for solutions of the Bogomolnyi equations
\be
\xi_i \mp  \frac{ig_2\sin^2 \xi}{(1+|u|)^2} \epsilon_{ijk} u_j \bar{u}_k =0.
\ee
Note, that this Bogomolnyi equation is identical to the one in the dilaton - $SU(2)$ Yang-Mills model describing a magnetic monopole \cite{Maison}-\cite{We-YM}. Further, it lead to the constraints (\ref{constr-2}), again. 

An example can be provided by a particular choice of $g_2$ and $h_2$,
\be
g_2=\frac{1}{\cos^2 \frac{\xi}{2}}, \;\;\; h_2=1 .
\ee
Then, using the ansatz in spherical polar coordinates, again, the topologically nontrivial solutions are again (\ref{con-sing})  for $u$, whereas 
 $\xi=\xi(r)$ obeys
\be
\frac{d\xi}{dr}=\pm \frac{2m}{r^2}{\sin^2 \frac{\xi}{2}} .
\ee
Imposing the following boundary conditions: $\xi(r=0)=0$ and $\xi(r=\infty)=\pi$, the pertinent solution is
\be
\cot \frac{\xi}{2} =  \frac{m}{r} \;\;\; \Rightarrow \;\;\; \xi = 2\mbox{ arccot }\frac{m}{r} .
\ee
Here,
\be
\langle g_2h_2 \rangle _{\mathbb{S}^3} = \frac{2}{\pi} \int_0^\pi d\xi \sin^2 \xi \cdot \frac{1}{\cos^2 \frac{\xi}{2}} =4
\ee
and the energy and topological charge are
\be
\tilde E^{(2)}=8\pi^2 m, \;\;\; B=m.
\ee

\subsection{The holomorphic map like model}

Now, we define the model
\be
\tilde{\mathcal{L}}^{(1)}=h_1 \left( \frac{u_\mu \bar{u}^\mu}{(1+|u|^2)^2}+g^2_1 \sin^4\xi \frac{ (\xi_\mu u^\mu)(\xi_\nu \bar{u}^\nu)  - \xi_\mu^2 (u_\nu\bar{u}^\nu)}{(1+|u|^2)^2} \right) .
\ee
The energy of the static case reads 
\bea
\tilde{E}^{(1)}&=& \int d^3 x h_1 \left( \frac{u_i  \bar{u}_i}{(1+|u|^2)^2} +g^2_1  \frac{\sin^4 \xi}{(1+|u|^2)^2} (i\epsilon_{ijk} \xi_j u_k) (-i\epsilon_{imn} \xi_m \bar{u}_n)  \right) \\
&=& \int d^3 x \frac{h_1}{(1+|u|^2)^2} (u_i\pm ig_1 \sin^2 \xi \epsilon_{ijk} \xi_j u_k)(\bar{u}_i 
\mp ig_1 \sin^2 \xi \epsilon_{imn} \xi_m \bar{u}_n) 
\nonumber \\ && \hspace*{1cm} \mp 2 \int d^3 x  \frac{ig_1h_1\sin^2 \xi}{(1+|u|^2)^2} \epsilon_{ijk} \xi_i u_j \bar{u}_k \nonumber \\
&\geq & 2 \left| \int d^3 x  \frac{ig_1 h_1 \sin^2 \xi}{(1+|u|^2)^2} \epsilon_{ijk}  \xi_i u_j\bar{u}_k \right| \nonumber\\
&=& 4\pi^2 \left| \int d^3x g_1 h_1 B_0 \right| = 4\pi^2 |B| \left\langle g_1h_1 \right\rangle _{\mathbb{S}^3} \nonumber
\eea
where the bound is saturated for solutions of the following Bogomolnyi equation
\be
u_i\pm ig_1 \sin^2 \xi \epsilon_{ijk} \xi_j u_k=0
\ee
and its complex conjugation. Note that this formula implies the constraints (\ref{constr-1}) for the complex and real fields. 
An example of BPS Skyrmions of this type can be found for 
\be
h_1=\sin^2\xi, \;\;\; g_1=\frac{1}{ \sin^3 \xi} .
\ee
We assume that $\xi=\xi (z)$ and $u=u(x,y)$ in cartesian coordinates $(x,y,z)$. Then, the scalar obeys
\be
g_2 \sin^2 \xi \xi_z=-1 \;\;\; \Rightarrow \;\;\; \xi_z=-\sin \xi
\ee
which is the sine-Gordon kink equation and therefore
\be
\xi = 2\arctan e^{-z} .
\ee
This means that the complex scalar obeys
\be
u_i \mp i\epsilon_{ij}  u_j=0
\ee
where $(i,j) \in \{ 1,2 \}$ and therefore it is just a holomorphic (anti-holomorphic) function in $x+iy$. For example, $u=\rho^m e^{im\varphi}$, when cylindrical coordinates $(\rho, \varphi ,z)$ are used. So, finally we get holomorphic 2-dimensional solitons located on a sine-Gordon brane with co-dimension one, and
\be
\tilde{E}^{(1)}=4\pi^2|B|\left\langle \frac{1}{\sin \xi} \right\rangle _{\mathbb{S}^3} = 16\pi |B|, \;\;\; B=m.
\ee

\section{Summary}
We found the interesting result that the generalized Skyrme model can, in fact,  be expressed as a sum of three BPS submodels,
\bea
\mathcal{L}_{Sk}&=&\mathcal{L}_2+\mathcal{L}_4+\mathcal{L}_6+\mathcal{L}_0 \nonumber \\
&=& \left( \mathcal{L}_2^{(1)}+\mathcal{L}_4^{(1)} \right) + \left( \mathcal{L}_2^{(2)}+\mathcal{L}_4^{(2)}\right) + \left( \mathcal{L}_6+\mathcal{L}_0 \right) \nonumber \\
&\equiv& \mathcal{L}^{(1)}_{BPS} + \mathcal{L}^{(2)}_{BPS} +\mathcal{L}_{BPS} . \label{gen-Sk}
\eea
In comparison with the BPS Skyrme model, however, the new BPS submodels reveal some important differences.
\begin{enumerate} 

\item The new BPS submodels cannot be obtained as a certain limit (particular values of the model parameters) of the full Skyrme model. Each of them consists of two terms - one emerging from the Dirichlet part and one from the Skyrme part. 

\item The new BPS submodels are completely independent of the pion mass.  The potential part of the generalized Skyrme model only contributes to the BPS Skyrme model.

\item While the BPS Skyrme model possesses well-behaved topologically nontrivial solutions i.e., BPS Skyrmions, 
whose compact or non-compact nature depends on the potential,
the other BPS submodels provide solutions with different characteristics. The second BPS submodel leads to fractional Skyrmions, i.e., to solutions with a non-integer baryon number. The first BPS submodel has compact Skyrmions as solutions. Obviously, the two submodels cannot have common solutions, because this would lead to a BPS solution for the full standard Skyrme model $\mathcal{L}_2 + \mathcal{L}_4$, which are known not to exist.

\item One characteristic feature of the solutions of the two submodels is the opposing effect which they have on the Skyrmion size. Indeed, the model $\mathcal{L}^{(1)}$ leads to finite size Skyrmions (compactons) whose size, in addition, is independent of the baryon number.   The model $\mathcal{L}^{(2)}$, on the other hand, leads to fractional Skyrmions, i.e., "Skyrmions" of "more than infinite size" which do not even fit in the infinite interval $r\in [0,\infty ]$. The size of the Skyrmions of the full model $\mathcal{L}_2 + \mathcal{L}_4$ is a compromise between these two extremes, i.e, Skyrmions with infinite size, which decay algebraically for $r\to \infty$.

\item  The submodel $\mathcal{L}^{(1)}$ leads to Skyrmions with a shell-like structure, where the energy density is zero both at the center $r=0$ and outside the compacton boundary and takes its maximum value at some nonzero radius. This reflects a known behaviour of the Skyrmions of the full   model $\mathcal{L}_2 + \mathcal{L}_4$, which have a shell-like structure, as well \cite{shells}. 

\end{enumerate}

As mentioned already, the submodel $\mathcal{L}^{(1)}$ is solved by arbitrary rational maps $u(z)$ for an ansatz $\xi (r)$ and $u(z)$ in spherical polar coordinates, and the same "ansatz" provides rather good approximations for solutions of the full standard Skyrme model $\mathcal{L}_2 + \mathcal{L}_4$. Here we just want to remark that inserting the rational map ansatz into the static energy functional of the generalised Skyrme model (\ref{gen-Sk}) leads exactly to the same restricted variational problem for the rational map $u=R(z)$. That is to say, the restricted energy functional is
\be
E_{\rm gen} = 4\pi \int dr \left[ r^2 \xi'^2 + \mu^2 r^2 \mathcal{U} (\xi) + 2 B \sin^2 \xi (\xi'^2 +1) +\mathcal{I} \frac{\sin^4 \xi}{r^2} \left(1+\frac{\lambda^2}{4}\xi^2 \right) \right]
\ee
where the rational map $R(z)$ must minimize the functional \cite{rat-map}
\be
\mathcal{I} = \frac{1}{4\pi} \int d\Omega_{\mathbb{S}^2} \left( \frac{1+z\bar z}{1+R\bar R}\right)^4 \left( R_z \bar R_{\bar z}\right)^2 
\ee
exactly as in the rational map approximation for the model $\mathcal{L}_2 + \mathcal{L}_4$. This fact was pointed out recently in \cite{Bjarke2} and used there for a detailed study of the $B=4$ Skyrmion (the helium nucleus) within the generalised Skyrme model.

In addition to introducing the proper BPS submodels of the standard (massless) Skyrme model, we also considered some generalisations, where each term in the submodels is multiplied by a field-dependent coupling function. If the coupling functions only depend on the profile function $\xi$, then the generalisations based on $\mathcal{L}^{(1)}$ continue to be of the holomorphic type, i.e., the $u$ field has a CP(1)-model energy density in a separation-of-variable ansatz. The second model, $\mathcal{L}^{(2)}$, generalises to a field theory which shares its
Bogomolnyi equations with the ones in the dilaton - $SU(2)$ Yang-Mills model for a magnetic monopole \cite{Maison}-\cite{We-YM}, where $\xi$ plays the role of the dilaton. Both generalised models lead to genuine topological solitons obeying the required boundary conditions for appropriate choices of the coupling functions.

There exists a lower dimensional counterpart of the Skyrme model, the baby Skyrme model \cite{baby} (here in the complex field formulation)
\be
\mathcal{L}_{baby}= \lambda_2 \mathcal{L}_{2} + \lambda_4 \mathcal{L}_4+\lambda_0\mathcal{L}_0=  \lambda_2 \frac{u_\mu \bar{u}^\mu}{(1+|u|^2)^2} - \lambda_4  \frac{(u_\mu \bar{u}^\mu)^2-u_\mu^2 \bar{u}^2_\nu }{(1+|u|^2)^4} - \lambda_0 \mathcal{U}
\ee
where one can distinguish two BPS submodels. The $CP^1$ (sigma) model
\be
\mathcal{L}_{CP^1} \equiv \mathcal{L}_{2}=\frac{u_\mu \bar{u}^\mu}{(1+|u|^2)^2}
\ee
and the baby BPS Skyrme model \cite{babyBPS}
\be
\mathcal{L}_{BPS} \equiv \mathcal{L}_4+ \lambda_0 \mathcal{L}_0 = \frac{(u_\mu \bar{u}^\mu)^2-u_\mu^2 \bar{u}^2_\nu }{(1+|u|^2)^4} - \lambda_0 \mathcal{U} .
\ee
Both BPS submodels are genuine baby Skyrme models. Further, they are true BPS theories, i.e., there exist Bogomolnyi equations and solutions which saturate a topological bound involving a topological index. These equations are 
\be
u_i \pm i \epsilon_{ij} u_j=0
\ee
for the CP(1) model and 
\be
 \frac{i\epsilon_{ij} u_i \bar{u}_j}{(1+|u|^2)^2}  \pm \sqrt{\lambda_0 \mathcal{U}}=0
\ee
for the BPS baby Skyrme model. Hence, we recognize a similar pattern to the one found in the $(3+1)$-dimensional Skyrme model. 

Let us remark that there exist further possibilities to find BPS  versions of the Skyrme model. One particular example, based on the same field contents but a different Lagrangian was constructed in \cite{Fer-Zak}. 
If the topology of the base space manifold is changed and allows to define additional topological indices, then further BPS sectors of the Skyrme model related to these new topological indices may be found \cite{manton}, \cite{can-zan}.
Another option which requires, however, to change the field contents by adding an infinite number of vector mesons was proposed and developed in \cite{SutBPS}, \cite{rho1}.

Finally, the hidden BPS structure of the standard Skyrme model revealed here might be related to possible supersymmetric versions of the theory \cite{jose}, \cite{nitta}, because there is a close relation between BPS sectors and supersymmetry, in general (see \cite{susy-baby} for the baby Skyrme model).

\section*{Acknowledgements}
The authors acknowledge financial support from the Ministry of Education, Culture, and Sports, Spain (Grant No. FPA 2014-58-293-C2-1-P), the Xunta de Galicia (Grant No. INCITE09.296.035PR and Conselleria de Educacion), the Spanish Consolider-Ingenio 2010 Programme CPAN (CSD2007-00042), and FEDER. AW was supported by NCN grant 2012/06/A/ST2/00396.

\end{document}